\documentclass[runningheads]{llncs}

\usepackage[T1]{fontenc}
\usepackage{graphicx}
\usepackage{todonotes}
\setuptodonotes{inline}
\usepackage{csquotes}
\usepackage[hidelinks]{hyperref}

\begin{document}


%
\title{The Silent Scientist} 

\subtitle{When Software Research Fails to Reach Its Audience}

\author{Marvin Wyrich\inst{1} \and
Christof Tinnes\inst{1,2} \and
Sebastian Baltes\inst{3} \and
Sven Apel\inst{1}
}
\authorrunning{Wyrich et al.}
%
\institute{Saarland University, Saarland Informatics Campus, Germany\\
\email{wyrich@cs.uni-saarland.de}
\and
Siemens AG, Germany\\
\and
University of Bayreuth, Germany}
\maketitle              

\noindent If software research were a performance, it would be a thoughtful theater play---full of rich content but confined to the traditional stage of academic publishing. 
Meanwhile, its potential audience is immersed in engaging on-demand experiences, leaving the theater half-empty, and the research findings lost in the wings.
As long as this remains the case, discussions about research relevance and impact lack meaningful context.

\section*{An identity crisis}
For researchers, reflecting on their own work is part of good scientific practice. This is common across all scientific disciplines, including software research.\footnote{By \enquote{software research,} we refer to the broader research field that includes software engineering and related disciplines focused on the theory and practice of building and maintaining software systems. We consider this field as a case study within computer science, where much of the research is application-oriented and should therefore have a relatively receptive target audience. Similar calls to improve research communication can be found in related fields, such as in information systems, where recent discussions have highlighted the potential of citizen science~\cite{Weinhardt:2020:citizen}.} Though critical self-reflection is a staple in research, the software research community seems to take it further, grappling almost yearly with feelings of insignificance and entirely questioning the relevance of their work. In the past two years, such discussions have been nearly impossible to track. Two prominent examples illustrate this: In 2022, Lionel Briand delivered a keynote at ICSE, the largest software engineering conference, critically discussing the impact we can expect from software research.\footnote{Lionel C. Briand's \href{https://conf.researchr.org/details/icse-2022/icse-2022-keynotes/4/Mathematicians-Social-Scientists-or-Engineers-The-Split-Minds-of-Software-Engineer}{ICSE 2022 keynote on \textit{the split minds of software engineering researchers}}.} Two years later, one of the leading software journals, the Journal of Systems and Software, launched a new column inviting software practitioners to share their perspective on making software research more relevant~\cite{Avgeriou:2024:DearResearchers}. Apparently, the software research community has reached a new peak in its ongoing quest to doubt its own relevance and impact.

To some extent, we understand the sentiment. Significant resources go into software research, to improve collaboration or drive innovation. It is reasonable to critically question how many of these research findings will be implemented in the near or distant future. In fact, considerable resources have already been devoted to addressing this very question, and analyses have identified concrete examples showing how software research contributed to advancements in development tools and methods, such as configuration management and programming languages.\footnote{See, for example, \href{https://www.nsf.gov/awardsearch/showAward?AWD_ID=0137766}{\textit{The Impact Project: Determining the Impact of Software Engineering Research Upon Practice}}, as well as international workshops on \href{https://insights.sei.cmu.edu/library/third-international-workshop-on-adoption-centric-software-engineering/}{\textit{adoption-centric software engineering}}.}

However, in the ongoing discussions surrounding the relevance and impact of software research, there is a notable oversight: the crucial role of \emph{science communication}.
It is surprising how often this aspect is disregarded.
Like a \emph{silent scientist}, quietly publishing papers and expecting their work to speak for itself, many researchers assume their findings will naturally find their audience.
This assumption can lead to an overly self-critical misconception: that if research lacks impact, it must be irrelevant.
In this essay, we set out to challenge this view.
Research can only have impact if it reaches its target audience in the first place, which requires making findings accessible through various communication channels~\cite{Nisbet:2009:SciCom,Wyrich:2023:TeachingSciCom}.
The tricky thing about this situation is that, so far, active science communication has been sparse in the area of software research, and those who have tried often find their efforts unrewarded or unsuccessful~\cite{Wilson:2024:NeverWorkTheory}.

\section*{Understanding the relevance of software research}
To appreciate the relevance of software research, we need to consider how different topics resonate with different stakeholder groups---what's impactful to one may be irrelevant to another.

First, \emph{software research} covers diverse content, which can be roughly divided into technical advancement and (human-centered) empirical understanding.
It seeks technical improvements, such as methods to automatically find and fix bugs, and empirical insights, such as understanding what factors influence the productivity of software developers.
Corresponding studies of the two types of research use entirely different research methodologies, with some focusing on technical evaluations without involving humans, while others include human participants as primary subjects.
This alone can create varied perceptions of relevance, as some studies directly involve the people they aim to help.
Additionally, measuring impact and implementing findings differ significantly.
For technological progress, impact is measured through adoption or metrics on quality and performance. Findings on practitioner collaboration require integration into sociotechnical processes, with impact measured through changes in behavior or improved satisfaction.

Second, different phases of the research process are of interest to different stakeholders.
A research project can span several years and progress through various phases, at the end of which one or more publications may appear. Judging the relevance of research based on a single publication inevitably leads to a large proportion of the readership not (yet) feeling addressed. Let us assume that we are looking for a source code quality metric that indicates the comprehensibility of source code. Thanks to software research, we know that most code comprehensibility metrics do not, in practice, reflect what they are supposed to measure~\cite{DBLP:conf/icse/PeitekAPBS21,Scalabrino:2019:Automatically}. 
So, we are well advised not to repeat the mistake and simply design another metric based on our gut feeling. We start a little earlier, define the notion of code comprehension, conduct basic research, and understand the neuropsychological correlates of code comprehension in the brain of developers to develop a meaningful metric in a subsequent step~\cite{Siegmund:2020:Crazy}. Will that conceptual definition of code comprehension or neurophysiological lab studies with small code snippets hold actionable insights for practice? Probably not. Are these studies relevant at all? Absolutely, because software research has more than just the target group of software practitioners. In this case, these early findings may help educators better support novice programmers with code comprehension difficulties, guide experimenters to design better studies and ask more targeted questions, and they eventually aid other researchers build their application-oriented research on a solid theoretical foundation.

The point is that relevance is much more complex than is often portrayed. What is irrelevant to one person today may be very relevant for another tomorrow. We assure the reader that this is not just a convenient excuse that allows researchers to retreat into an ivory tower. We ourselves see systemic difficulties that reinforce the impression of a lack of relevance of software research. For example, the very people we want to help with our research at any given time are often involved too late in the research process. 
Researchers who have not spoken and, in particular, \emph{listened} to their target group risk creating an artificially constructed problem space.  Listening can take many forms, such as attending industry meetups, conducting targeted surveys, or monitoring developer forums and social media. Without these insights, it becomes difficult to argue for the relevance of one's research, and even harder to attract anyone as an audience for science communication. 
Another challenge is certainly that software researchers are incentivized by the scientific review and funding process to present solutions to be as generalizable as possible. As a result, we hear of software practitioners who fail to adapt published research findings because research has raised false expectations of applicability.

Note that, taking this viewpoint, we will neither succeed in fundamentally rethinking the academic system nor in convincing every reader of the relevance of every single research paper. We do not even want to. There are already enough opinions on why some software research is theoretical in nature and why other research should be more application-oriented. There is a multitude of opinions on what role industry--academia collaborations could play for the relevance of research topics. And there are certainly enough opinions on how the impact of software research could be evaluated. Our point is different: \emph{As long as the research community does not manage to make its research accessible to the respective target group at all, we do not even need to talk about relevance and impact.}

\section*{No impact without science communication}
The feeling of irrelevance and missing impact can often be traced back to a communication problem: The reality is that nobody cares about your research unless you make them care. This does not happen by itself with the publication of a research paper. The notion that software practitioners will dive into the academic world, eagerly browse online libraries for papers, and read them with sparkling eyes is as romantic as it is unlikely. A far more probable scenario for success is when researchers and practitioners build a bridge between these two worlds through actively engaging in dialogue.

The fact that software practitioners are indeed interested in contemporary software research is shown by a recent study that examines how these research findings are disseminated and discussed on LinkedIn~\cite{Wyrich:2024:LinkedIn}. 
These data reveal that the majority of individuals who post content and comment on the posts are software practitioners.
The authors of that study conclude that researchers are not doing enough themselves. They note that some software research is so engaging that practitioners take on the role of science communicators---a role traditionally reserved for researchers, but inadequately filled by them in the case of software research~\cite{Wyrich:2024:LinkedIn}.

The good news is that researchers can do a lot to improve science communication without having to revamp the publication system or redesign research processes.
If we follow Burns et al.~\cite{Burns:2003:SciComDefinition} in their definition of science communication \enquote{as the use of appropriate skills, media, activities, and dialogue to produce one or more of the following personal responses to science: Awareness, Enjoyment, Interest, Opinion-forming, and Understanding}, it is difficult to find a software research paper that cannot be brought to the target group with, at least, one of these intentions. Therefore, post-publication science communication is something every researcher can and should engage in.

Why isn't it happening?
We see two main reasons. First, publications are the currency of academia. Career advancement largely depends on the number of top-tier publications, while outreach and the practical impact of research are often secondary in university selection processes. Considering such systemic incentives, it is understandable that researchers quickly move on to the next project after publishing.\\ 
Second, there is a lack of evidence on the effectiveness of science communication in software research, so it may be useful to look at other sciences. Bauer et al.'s introduction to implementation science, i.e., \enquote{a science of implementation}, is only 10 years old, but has since been cited over 2000 times, many times by successful field reports~\cite{Bauer:2015:Introduction}. In the context of clinical research, \enquote{the relatively new field of implementation science has developed to enhance the uptake of evidence-based practices and thereby increase their public health impact}~\cite{Bauer:2020:ImplementationScience}. Software research needs similar initiatives to learn how to bring research findings into software practice. In contrast, empirical insights on how research transfer works in software engineering and how science communication actually affects software engineering practice are still missing. As long as this gap persists, it is challenging to convince researchers that the extra effort is worthwhile.

In any case, what is hard to dispute is that research is unlikely to have much impact without \emph{any} communication.
This can take the form of a blog post, a social media discussion, a workshop or Dagstuhl seminar involving the target audience, or any other channel that fits the audience~\cite{Cooke:2017:Considerations,Illingworth:2017:DeliveringEffective}.
ACM itself offers several channels for this very purpose, such as the CACM blog and the ACM Queue.
Even something as simple as submitting a ticket in an open-source project that you have studied in your research can help connect your work to those who would benefit. 
While the software research community may not yet know whether these efforts will lead to new collaborations, more citations, or industry-wide change, there is reason to assume that effective communication can make a difference.
To provide, at least, anecdotal encouragement, we quote software researcher Marcos Kalinowski, who wrote on social media~\cite{MarcosTweet}: \enquote{Recently I shared the result of a PhD thesis on LinkedIn and it reached 4,000+ reactions and 270,000+ impressions. 95\% of my network is from industry. We are well equipped to burst the academic bubble!}

Looking ahead, researchers are well advised to ensure their work reaches the right audience by complementing traditional, high-quality academic paper publishing with efforts to make research more accessible to relevant stakeholders. Identify your target audience to tailor your message! Use diverse communication channels beyond papers, and actively engage with practitioners to foster dialogue rather than broadcasting information! This approach can spark feedback, opening doors to new ideas and collaborations that shape future research.

If you are in software research, or any other research area for that matter, remember why you began your research journey. The \emph{silent scientist} may publish papers, but without reaching the right audience, this work risks going unnoticed and unappreciated. We therefore urge researchers to break the silence and actively approach those communities that stand to benefit most from the findings.

\section*{Acknowledgments}
We thank two reviewers and the associate editor for their constructive comments, which helped to enrich the article with additional perspectives. This work has been supported by the European Union under ERC Advanced Grant ``Brains On Code'' (101052182).

\section*{Author Information}
Marvin Wyrich is a Postdoctoral Researcher at the Saarland University, Saarland Informatics Campus, Germany.
Christof Tinnes is a Senior Key Expert at Siemens, Germany, and a PhD student at Saarland University, Germany.
Sebastian Baltes is a Professor of Software Engineering at the University of Bayreuth, Germany.
Sven Apel is a Professor of Software Engineering at the Saarland University, Saarland Informatics Campus, Germany.

%
%
\bibliographystyle{splncs04}
\bibliography{main}

\end{document}